# Docking study for Protein Nsp-12 of SARS-CoV with Betalains and Alfa-Bisabolol.

Isaac Lucas-Gómez[1,2], López-Fernández Abelardo[1], González-Pérez Brenda Karen[1,3], Rivas-Castillo Andrea M.[1], Valdez Calderón A.[1] & Gayosso-Morales Manuel A.[1]

1. Universidad Tecnológica de la Zona Metropolitana del Valle de México. C.P. 43800 Tizayuca, Tizayuca Hidalgo.

2. Doctorado en Nanociencias y Nanotecnología, Centro de Investigación y de Estudios Avanzados del Instituto Politécnico Nacional Ciudad de México. C.P. 07360.

3. Universidad Nacional Autónoma de México, Campus Iztacala. Av. De los Barrios #1. Col.Los Reyes Iztacala, Tlalnepantla, Estado de México C.P. 54090, México.

**Abstract:**

The present Health Crisis tests the response of modern science and medicine to finding treatment for a new COVID-19 disease. The presentation on the world stage of antivirals such as remdesivir, obeys to the continuous investigation of biologically active molecules with multiple theoretical, computational and experimental tools. Diseases such as COVID:19 remind us that research into active ingredients for therapeutic purposes should cover all available sources, such as plants. In the present work, in silico tools, specifically docking study, were used to evaluate the binding and inhibition capacity of an antiviral such as remdesivir on the NSP-12 protein of SARS-CoV, a polymerase that is key in the replication of the SARS-COV virus. The results are then compared with a docking analysis of two natural products (Alpha-Bisabolol and betalain) with SARS-CoV protein, in order to find more candidates for COVID-19 virus replication inhibitors. in addition to increasing studies that help explain the specific mechanisms of the SARs-CoV-2 virus, remembering that we will have to live with the virus for an indefinite time from now on. Finally, natural products such as betalains may have inhibitory effects of a small order but in conjunction with other synergistic active ingredients they may increase their inhibition effect on NSP-12 protein of SARS-CoV.

## 1. Introduction

The Coronaviruses (CoVs) represent a heterogeneous family of positive sense RNA viruses and they are able to cause respiratory and enteric diseases in human and animal hosts. About all the types of coronavirus that exist, the most relevant are the highly pathogenic human CoVs, Severe Acute Respiratory Syndrome (SARS-CoV) and Middle East Respiratory Syndrome (MERS-CoV) all of them are able to provoke serious respiratory diseases (1-4). The zoonotic SARS-CoV-2 emerged in 2019 where the first cases reported for the first time were in Wuhan (China) and spread throughout the world until to be declared a pandemic on March 11, 2020 (4-7). Viruses similar to SARS-CoV-2 (Covid-19) continue transiting in the reservoirs of bats and other species of mammals belonging to wildlife (4). Outbreaks of highly pathogenic human CoVs still being an emerging threat to global health security and are likely to continue to occur in the future.

Emerging viruses of the CoV family require special attention to develop antiviral strategies aimed at conserved elements of the viral life cycle, such as the viral mechanism responsible for replication and transcription of the viral positive-chain RNA genome (2, 4). The CoV RNA synthesis mechanisms, is of multiple subunit, it is a set of non-structural proteins (nsp) produced as breakdown products of the viral polyproteins ORF1a and ORF1ab5. The RNA-dependent RNA polymerase nsp12 has minimal activity by itself, only with the addition of the cofactors nsp7 and nsp8 is that they can stimulate the activity of the polymerase nsp12 (6). It is necessary extra viral nsp subunits for carrying out replication and transcription activities; however, the nsp12-nsp7-nsp8 complex nowadays represents the minimum required formulation for nucleotide polymerization (6).

After the emergence of SARS-CoV, there was an intense effort for structurally, characterize the CoV replication complexes. This EFFORT GAVE AS A result in the determination of the high-resolution structure for many of the SARS-CoV nsps THROUGH X-ray crystallography and nuclear magnetic resonance. Besides this, complexes of nsp7-nsp8 and nsp10-nsp14 were also determined (2, 6).

Then in 2019 the scientist Kirchdoerfer, Robert N., *et. al.* Determinated the structure of the SARS-CoV nsp12 complex linked to the cofactors nsp7 and nsp8 by using cryoelectronics microscopy (cryo-TEM). The complex of nsp12, nsp7 and nsp8 has 160 kDa and make up most of the viral RNA synthesis complex of coronaviruses (6).

It is necessary to emphasize that the SARS-CoV protein belongs to the SARS 2002 epidemiological outbreak and the Covid-19 disease has a SARS-CoV-2 protein, both proteins have huge similarities and they are highly conserved in both viruses.

Starting with the contribution of the structure of the SARS-CoV nsp12 complex linked to the cofactors nsp7 and nsp8 (protein best identified by the PDB protein database). It begins the possibility to search for the nsp12 inhibitory structures with molecular analysis tools (docking) in order to inhibit the biological machanism of synthesis of SARS-CoV and one of the most promising molecules to do that is the Remdesivir, which is one of the first prescription inhibitors in patients with Covid-19 in the United States of America (1, 2, 6, 8).

Remdesivir, is a nucleotide analog that mimics the structure of adenosine. Gilead Sciences, Inc. originally developed it. So as to treat Ebola (4, 7, 9, 10). Even though it has not passed the phase 3 of the clinical trial of Ebola treatment, it showed a promising moderate improvement over the death rate from deadly Ebola and the active setting of remdesivir found already hydrolyzed and decorated with triphosphates, by using the remdesivir nucleus as the nucleoside. We call this hydrolyzed and phosphorylated remdesivir "RemTP". Like other nucleotide analogs, remdesivir could potentially be used as a wide spectrum antiviral drug, due to the structural similarities of RNA polymerase of various viruses (7, 9). Because of this, the Food and Drug Administration (FDA) approved remdesivir on October 22, 2020 for its treatment in severely covid-19 patients; due to the significant reduction in the duration of invasive medical intervention in positive patients for covid-19 (1, 2, 4, 7).

The new and rapid development of COVID-19 around the world is going to address the urgent need for more information about remdesivir as an inhibitor, including the search for conserved interaction points (amino acids) in the SARS-COV-2 protein (4). Therefore, we did a physics-based molecular modeling study on bonding mechanism between remdesivir and SARS-CoV-2 by using molecular docking techniques to identify candidates for inhibitors of the SARS-COV complex, which has similarity to the SARS-CoV-2 protein (covid-19) (1, 2, 6). Within the study, was modeled the interaction of the hydrolyzed and phosphorylated remdesivir nucleoside RemTP against the SARS-CoV protein as well as, 2 extra molecules called natural products, in the search for more inhibitors of the SARS-CoV protein.

The two products selected for SARS-CoV docking analysis are Alpha-Bisabolol (major component of Matricaria chamomilla and its essential oil) (11) and betalain (member of a family of secondary nitrogenous metabolites of plants who has red and yellow pigments) (12).

## 2. Materials and methods.

Molecular docking study

ChemBioOffice constructed ligands and the energy minimization performed with Molecular Mechanic by ChemBioDraw ultra software. Gaussian 03W, GaussView 3.0 and Chimera 1.14, using a Hartree-Fock calculation for the molecule in neutral basal state (without electrical charge), achieved an approach to the real molecular spatial array. Once the calculation was completed, we saved the final molecular structure and information in mol2 format for later manipulation in UCSF Chimera 1.14 (13, 14). The 3D structure of SARS-CoV in PDB format was obtained from Protein Data Bank (PDB) with identification (ID) 6NUR, as well as the identification in the National Center for Biotechnology Information (NCBI) is: ID 31138817. Finally, after obtaining of ligands and target proteins, the docking procedure was performed and optimized by the SIB web site service (11, 15, 16), the docking full procedure was reported by *Ortiz, Mario I., et. al.* in 2016.

## 3. Results and Discussion.

Zhang, L., and Zhou, R. (2020) described the first interactions of RemTP with a modified version of 6NUR-SARS-CoV (updated to reflect changes in the amino acid sequence for the Covid-19 protein) (1, 2). They were identified, the main positions where the active form of Remdesivir binds: the adenosine Tri Phosphate (ATP) part interacts with positive charges in the residues: LYS-551, ARG-553, ARG-555, LYS-621, LYS- 798, ARG-836 and original remdesivir section interacts with SER-549, ARG-555, LYS-545, ALA-547 and VAL-557 (1, 2). There are two sections of interaction, that of ATP and that of the nucleoside analog of the original remdesivir. The interaction of remdesivir blocks the main function of the polymerase complex nsp12-nsp8 and nsp7 (1). In addition to the aforementioned residues, interactions of RemTP with ASP-618 has been reported, this indicates that there may be more interactions that require further study (1, 2), but in this work, we will contrast the interactions reported by Zhang, L., and Zhou, R., versus to the results of interactions of two natural products: Betaline and Alpha-Bisabolol.

### 3.1 Docking SARS-CoV with RemTP.

Key amino acid positions were identified and reported by Zhang, L., and Zhou, R. within the 3D model of SARS-CoV (1). Later, the interactions near the key positions were recognized (only the positions were taken into account, since that the amino acids between the SARS-CoV polymerase and the SARS-CoV-2 modification differ in amino acids), as well as the Gibbs free energy (ΔG) of all positions considered feasible.

The complete docking calculation generated 250 possible interactions, of which 65 positions selected due to their proximity to the reported positions; of the 65 positions, the eight most favored energetically are reported (table 1).

Table 1 Energetically favored interactions for SARS-CoV-RemTP

| Interaction | Gibbs free energy (ΔG) kcal / mol |
|---|---|
| 53 | -16.27044 |
| 70 | -15.392919 |
| 94 | -17.583681 |
| 110 | -16.438456 |
| 144 | -17.546692 |
| 183 | -16.441689 |
| 206 | -18.714552 |
| 229 | -17.919659 |

Having the main interactions of RemTP with its corresponding ΔG we can make a direct comparison with the results of the selected natural products.

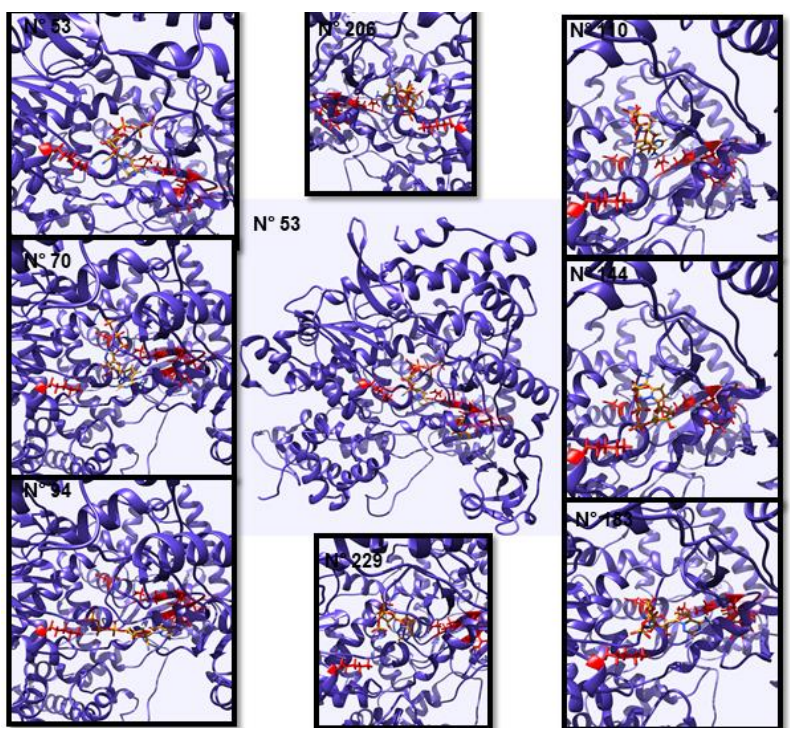

Figure 1. Eight main interactions (53, 70, 94,110,144,183,206 and 229) of RemTP (orange) with SARS-CoV. The main interactions of SARSCoV (purple) based on the work of Zhang, L., and Zhou, R. are: GLU-542, GLU-549, VAL-551, CYS 553, CYS-555, SER-556, LEU-557, TYR-558 and LYS 682, identified in red. Picture created with Chimera 1.14

### 3.2 Docking SARS-CoV with alfa-Bisabolol.

The complete docking calculation generated 250 possible interactions, of which 2 positions were selected due to their proximity to the positions reported in the literature, these most favored energy positions are reported in Table 2.

Table 2 Energy favored interactions for SARS-CoV-Alpha-Bisabolol.

| Interaction | Gibbs free energy (ΔG) kcal/mol |
|---|---|
| 143 | -6.545075 |
| 224 | -6.3719144 |

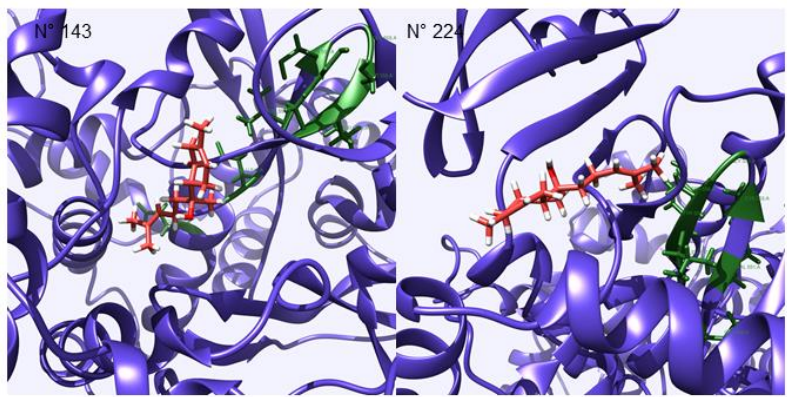

Figure 2. Two main interactions (143 and 224) of alpha-bisabolol (red) with SARS-CoV (purple). Picture created with Chimera 1.14.

### 3.3 Docking SARS-CoV with betalain.

The complete docking calcation generated 260 possible interactions, of which 49 have interaction with the sites reported in RemTP literature, of these 4 positions were selected for their most favored energy, they are reported in table 3.

Table 3. Energetically favored interactions for SARS-CoV-betalain.

| Interaction | Gibbs  Free Energy (ΔG) kcal/mol |
|---|---|
| 21 | -10.112583 |
| 47 | -7.7863874 |
| 59 | -9.799728 |
| 231 | -7.053678 |

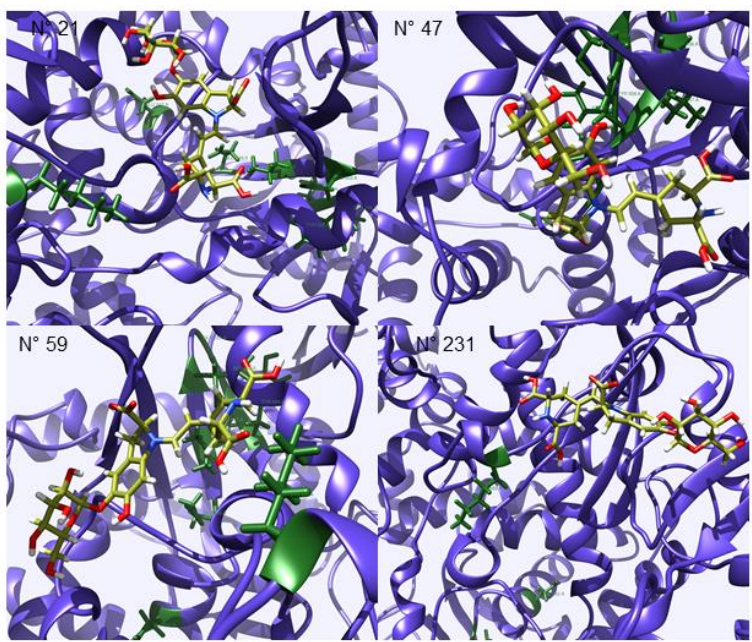

Figure 3. Four main interactions (21, 47, 59 and 231) of beta lain (yellow) with SARS-CoV (purple). Picture created with Chimera 1.14.

### 3.4 Discussion.

Analyzing the interactions between the three molecules, we observe that the largest number of interactions with the SARS-CoV protein is with RemTP, something already expected knowing that it is the active form of remdesivir, it is importan to emphasize that with 65 interactions and 8 favored energy positions the RemTP presents high binding affinity for SARS-CoV, this reinforces the inhibition studies of Zhang, L., and Zhou, R. as well as the extra information reported in the literature (2, 7). It is also so important to emphasiz that the molecules proposed as inhibitors must be structurally similar with nucleosides in order to guarantee a greater number of interactions and a high binding affinity with SARS-CoV, this is in order to keep looking to inhibit their RNA polymerase action. The presence of the amine functional group is important to increase interactions with positive partial charges of the SARS-CoV protein.

Based on the fact that the affinity for SARS-CoV is increased by the presence of the amine functional group, in the results we see the greatest interactions, within the two natural products it is betalain which has two amines in its structure. Betalain has 49 binding interactions and 4 energetically favored positions, being the lowest Gibbs free energy. This indicates that betalain could have an inhibitory effect on the SARS-CoV protein, although this is of low intensity when compared to the active form of remdesivir.

In the case of Alpha-Bisabolol, of the 250 positions generated by the docking study, only 2 turned out to have an interaction close to the active positions reported in the literature for SARS-CoV. It turns out, that Afa-Bisabolol is an example of low binding affinity with SARS-CoV and therefore, its inhibition capacity is null. However, a more in-depth analysis of natural products is required to identify structures with higher binding affinity.

## 4. Conclusions.

Docking **studies are a non-presential study to identify biologically active molecules towards a specific protein, through the elimination of molecules** with zero affinity and emphasizing those molecules to synthesize and test them experimentally. In the case of natural products, there are new molecules describe constantly in the scientific literature and whose interaction and biological effectiveness must be tested against proteins that represent problems of impact for society. In this case, the study demonstrates the feasibility of using the docking study to identify possible inhibitors of the SARS-CoV protein (key to the replication machinery of the virus that generates Covid-19).

Because of the natural products are chemical compounds of a varied chemical nature, they present a great possibility of having biological effects on SARS-CoV in addition to being compounds that were generated in living beings, they present a great capacity for degradation without becoming persistent pollutants in the environment. In the present work, we identify that the red pigment betalain mainly present in several fruits and vegetables is a candidate for inhibition of the SARS-CoV protein. It does not mean, that betalain is a complete treatment, but it could indicate the chemical structures with the greatest capacity of inhibition to find molecules more related to SARS-CoV and break the replication chain of the covid-19 virus.

## Acknowledgements.

Thanks to TSU Environmental Chemistry Technology (UTVAM) student Adriana Naxeli Pacheco Hernández for her help in data processing.

## 5. References

1.  Zhang L, Zhou R. Binding mechanism of remdesivir to SARS-CoV-2 RNA dependent RNA polymerase. 2020.
2.  Zhang L, Zhou R. Structural Basis of the Potential Binding Mechanism of Remdesivir to SARS-CoV-2 RNA-Dependent RNA Polymerase. The Journal of Physical Chemistry B. 2020;124(32):6955-62.
3.  Cohen J, Kupferschmidt K. Strategies shift as coronavirus pandemic looms. Science. 2020;367(6481):962.
4.  Kaye AD, Cornett EM, Brondeel KC, Lerner ZI, Knight HE, Erwin A, et al. Biology of COVID-19 and related viruses: epidemiology, signs, symptoms, diagnosis, and treatment: Considerations for Providing Safe Perioperative and Intensive Care in the Time of Crisis. Best Practice & Research Clinical Anaesthesiology. 2020.
5.  Organization WH. Coronavirus disease 2019 (COVID-19) Situation Report–52 URL: https://www. who. int/docs/default-source/coronaviruse/situationreports/20200312-sitrep-52-covid-19. pdf? sfvrsn= e2bfc9c0_4. Accessed on. 2020;12.
6.  Kirchdoerfer RN, Ward AB. Structure of the SARS-CoV nsp12 polymerase bound to nsp7 and nsp8 co-factors. Nature Communications. 2019;10(1):2342.
7.  Veronin MA, Lang A, Reinert JP. Remdesivir and Coronavirus Disease 2019 (COVID-19): Essential Questions and Answers for Pharmacists and Pharmacy Technicians. Journal of Pharmacy Technology. 2020:8755122520967634.
8.  Wang Y, Zhang D, Du G, Du R, Zhao J, Jin Y, et al. Remdesivir in adults with severe COVID-19: a randomised, double-blind, placebo-controlled, multicentre trial. Lancet. 2020;395(10236):1569-78.
9.  Tchesnokov EP, Feng JY, Porter DP, Götte M. Mechanism of inhibition of Ebola virus RNA-dependent RNA polymerase by remdesivir. Viruses. 2019;11(4):326.
10. Goldman JD, Lye DCB, Hui DS, Marks KM, Bruno R, Montejano R, et al. Remdesivir for 5 or
1   0 Days in Patients with Severe Covid-19. 2020;383(19):1827-37.
11. Ortiz MI, Fernández-Martínez E, Soria-Jasso LE, Lucas-Gómez I, Villagómez-Ibarra R, González-García MP, et al. Isolation, identification and molecular docking as cyclooxygenase (COX) inhibitors of the main constituents of Matricaria chamomilla L. extract and its synergistic interaction with diclofenac on nociception and gastric damage in rats. Biomedicine & Pharmacotherapy. 2016;78:248-56.

12. Lucas-Gómez I, Carrasco-Torres G, Bahena-Uribe D, Santoyo-Salazar J, Fernández-Martínez E, Sánchez-Crisóstomo I, et al. Green synthesis of silver nanoparticles with phytosterols and betalain pigments as reducing agents present in cactus Myrtillocactus geometrizans. MRS Advances. 2020:1-9.

13. Pettersen EF, Goddard TD, Huang CC, Couch GS, Greenblatt DM, Meng EC, et al. UCSF Chimera--a visualization system for exploratory research and analysis. Journal of computational chemistry. 2004;25(13):1605-12.

14. Huang CC, Meng EC, Morris JH, Pettersen EF, Ferrin TE. Enhancing UCSF Chimera through web services. Nucleic Acids Research. 2014;42(W1):W478-W84.

15. Grosdidier A, Zoete V, Michielin O. Fast docking using the CHARMM force field with EADock DSS. Journal of computational chemistry. 2011;32(10):2149-59.

16. Grosdidier A, Zoete V, Michielin O. SwissDock, a protein-small molecule docking web service based on EADock DSS. Nucleic Acids Res. 2011;39(Web Server issue):W270-7.

Supporting information

All picture created with Chimera 1.14.

RemTP interaction 53 with NSP-12 SARS-CoV

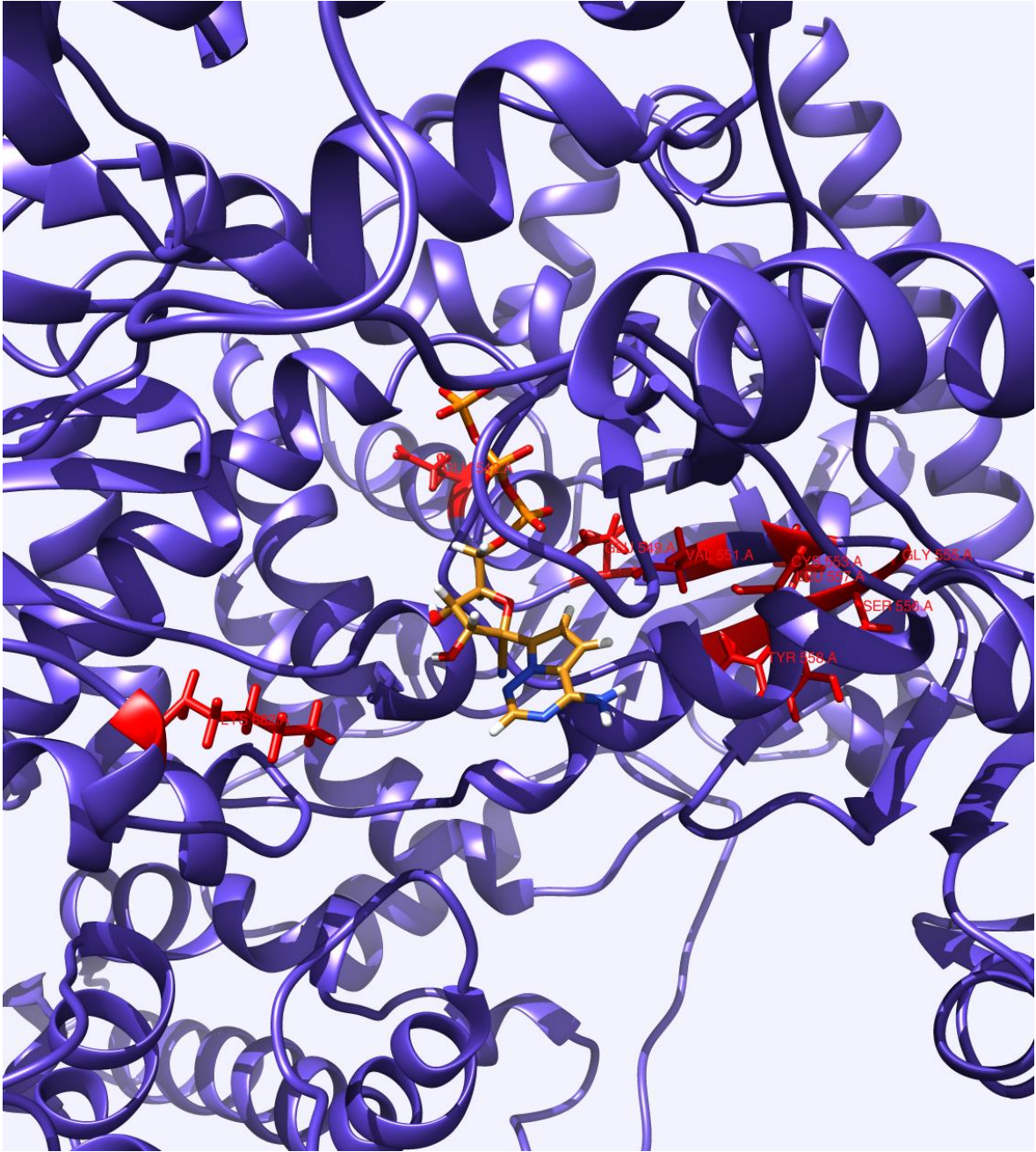

RemTP interaction 70 with NSP-12 SARS-CoV

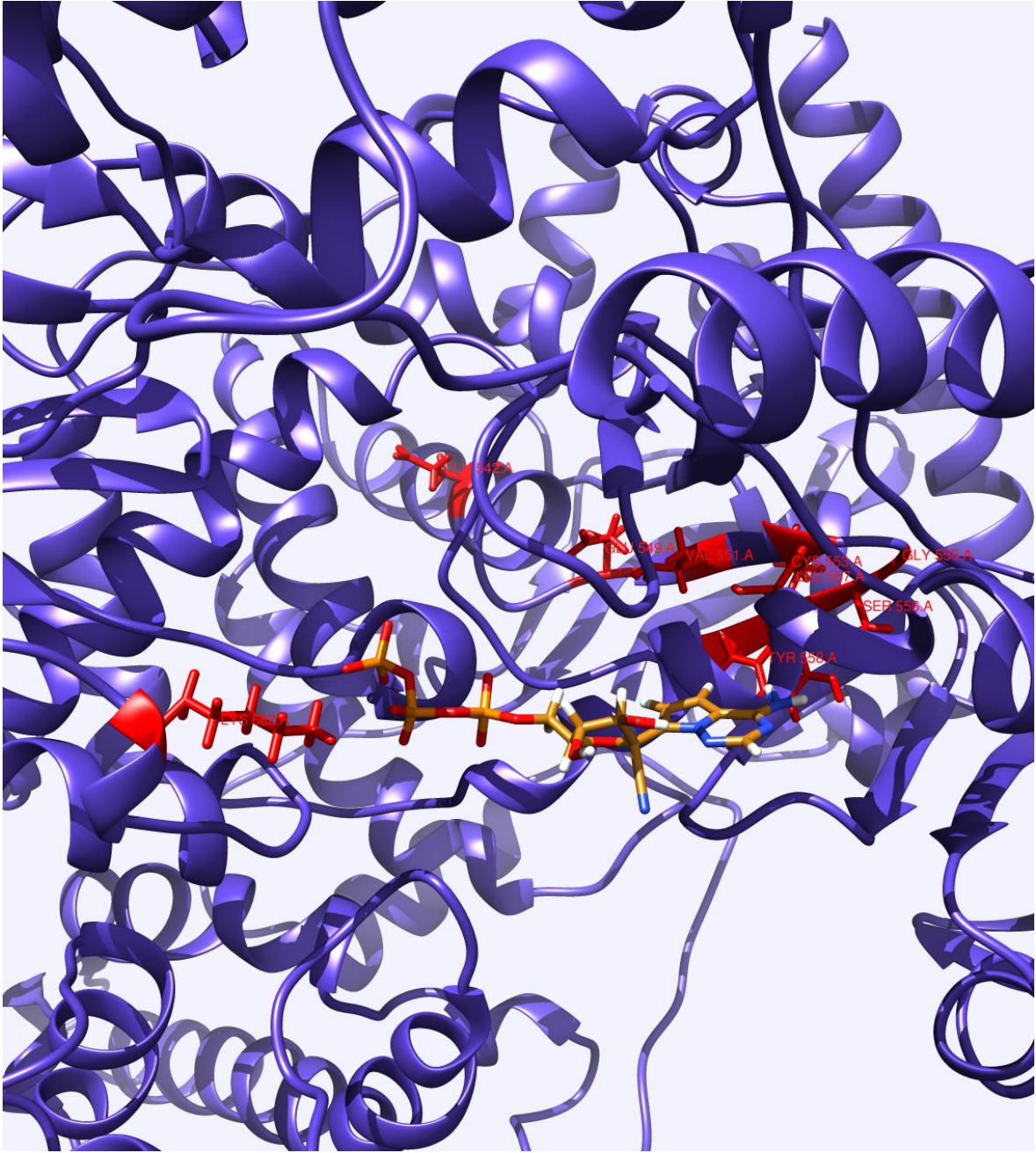

RemTP interaction 94 with NSP-12 SARS-CoV

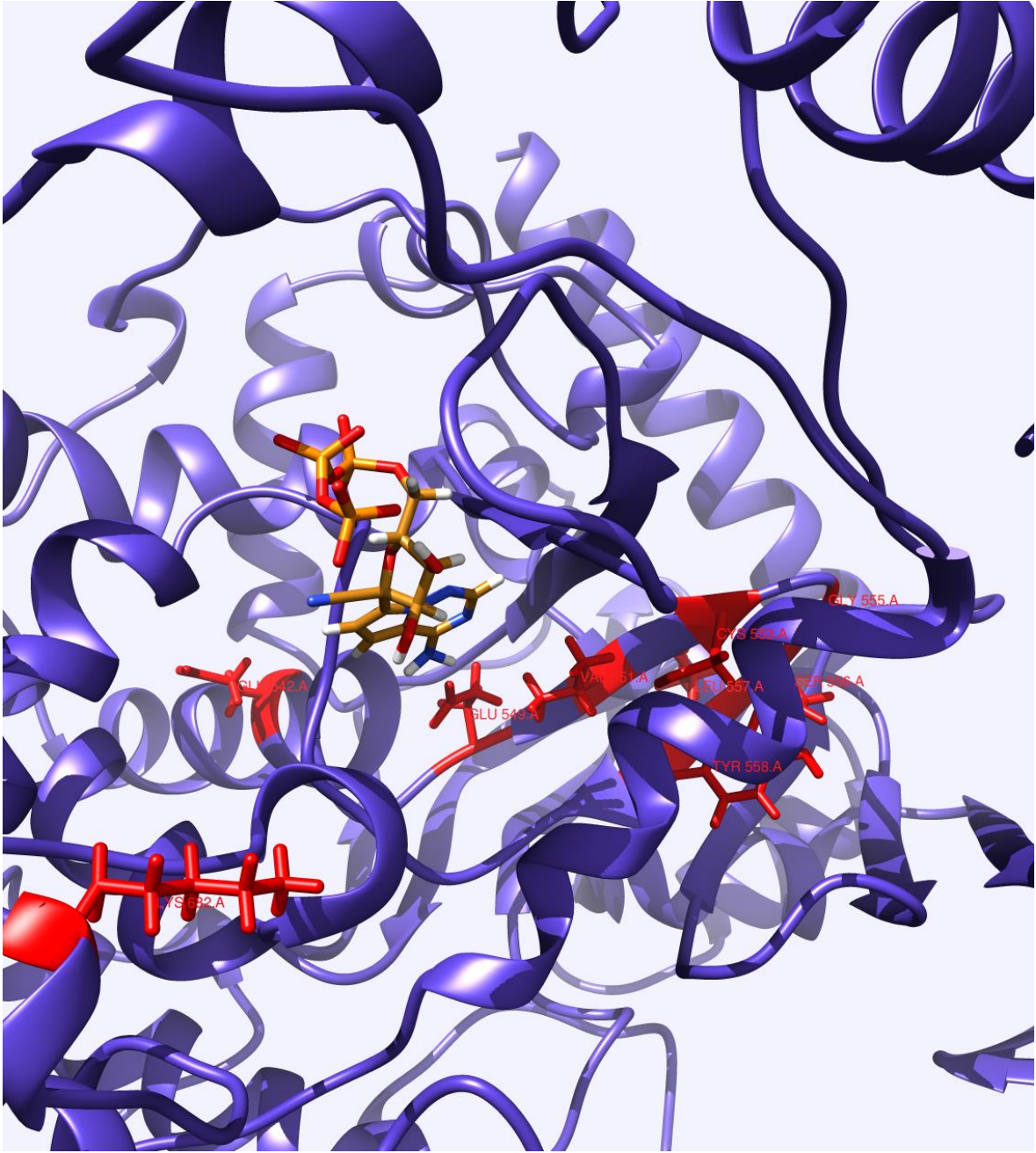

RemTP interaction 110 with NSP-12 SARS-CoV

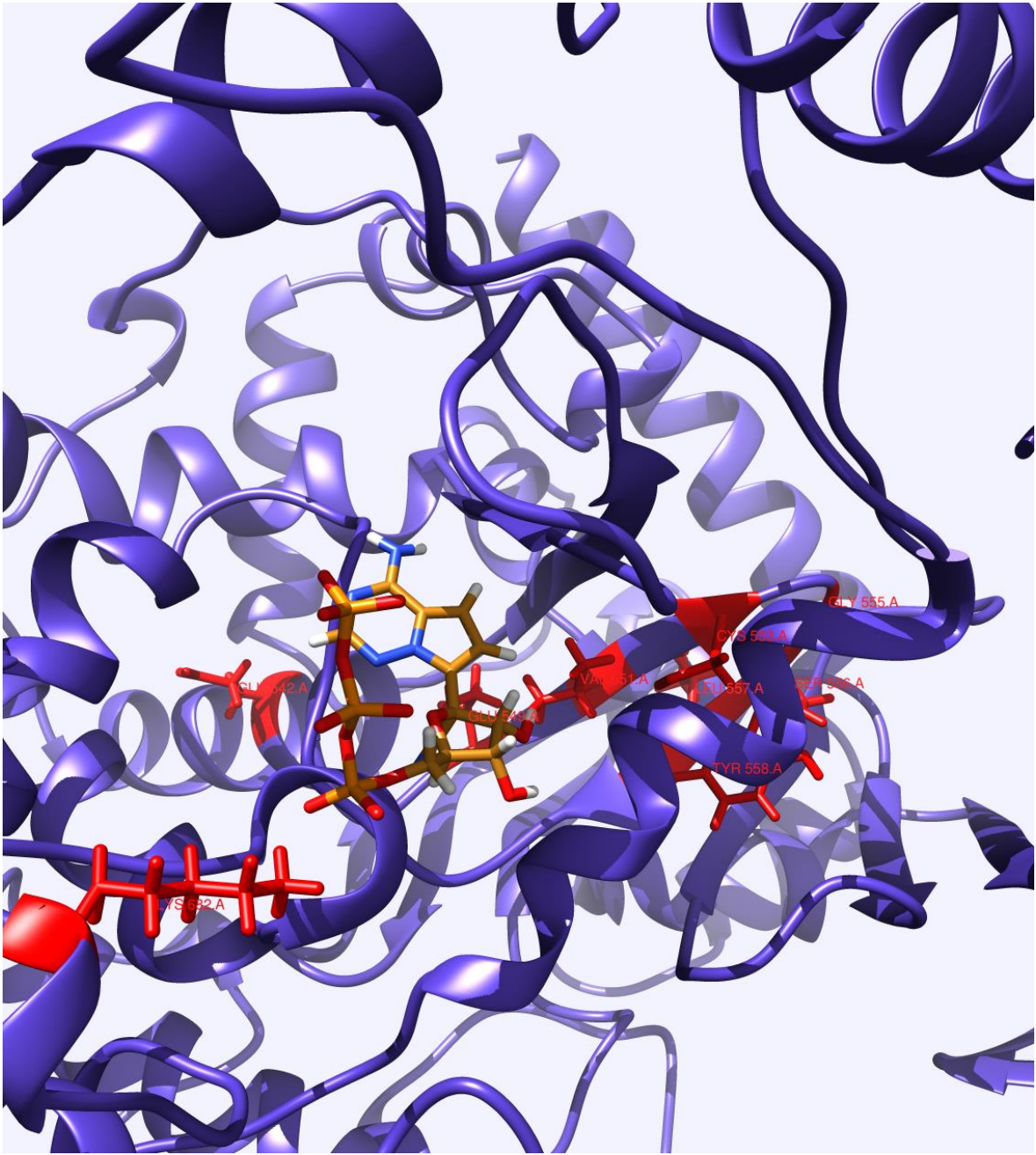

RemTP interaction 144 with NSP-12 SARS-CoV

RemTP interaction 183 with NSP-12 SARS-CoV

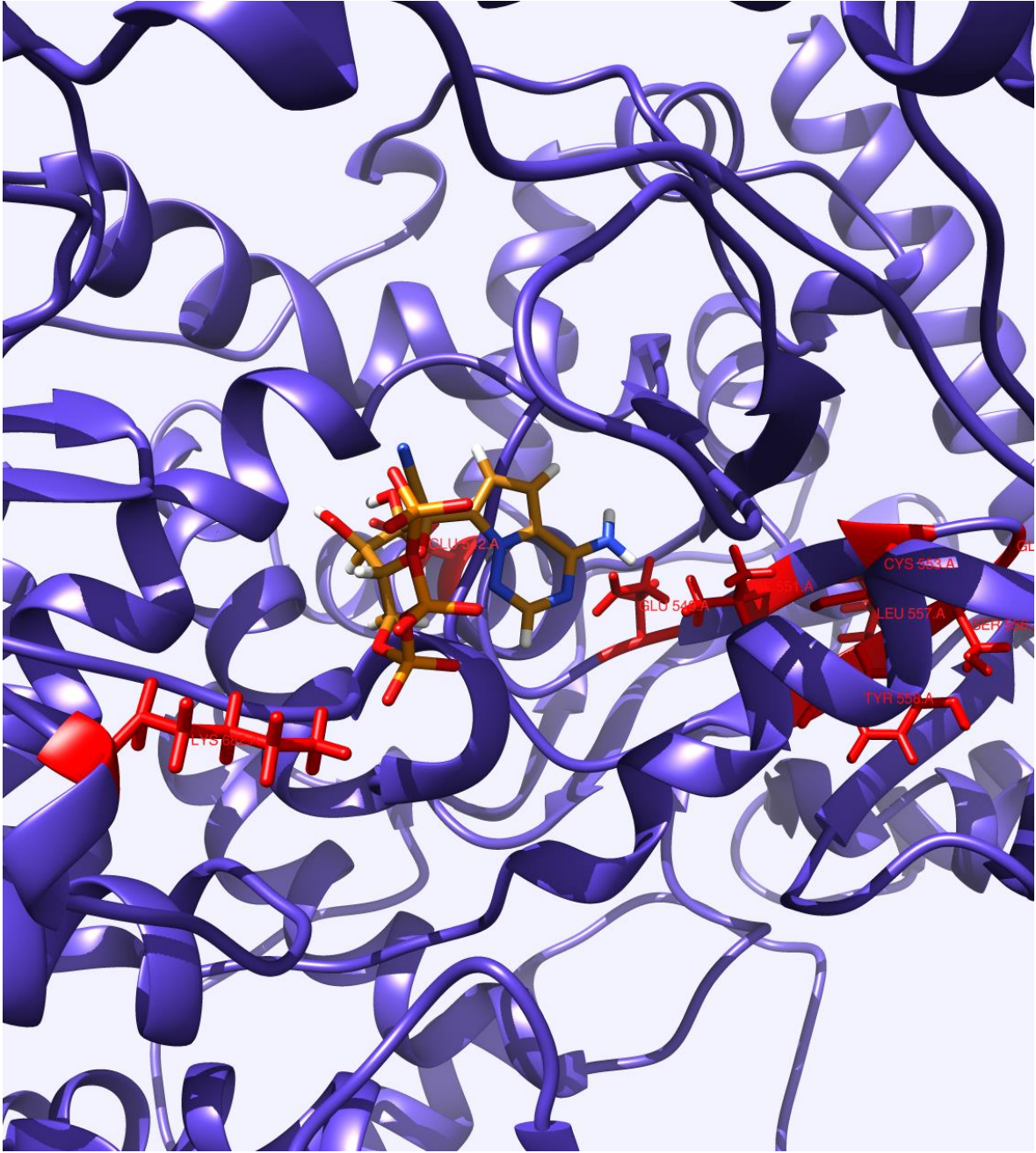

RemTP interaction 206 with NSP-12 SARS-CoV

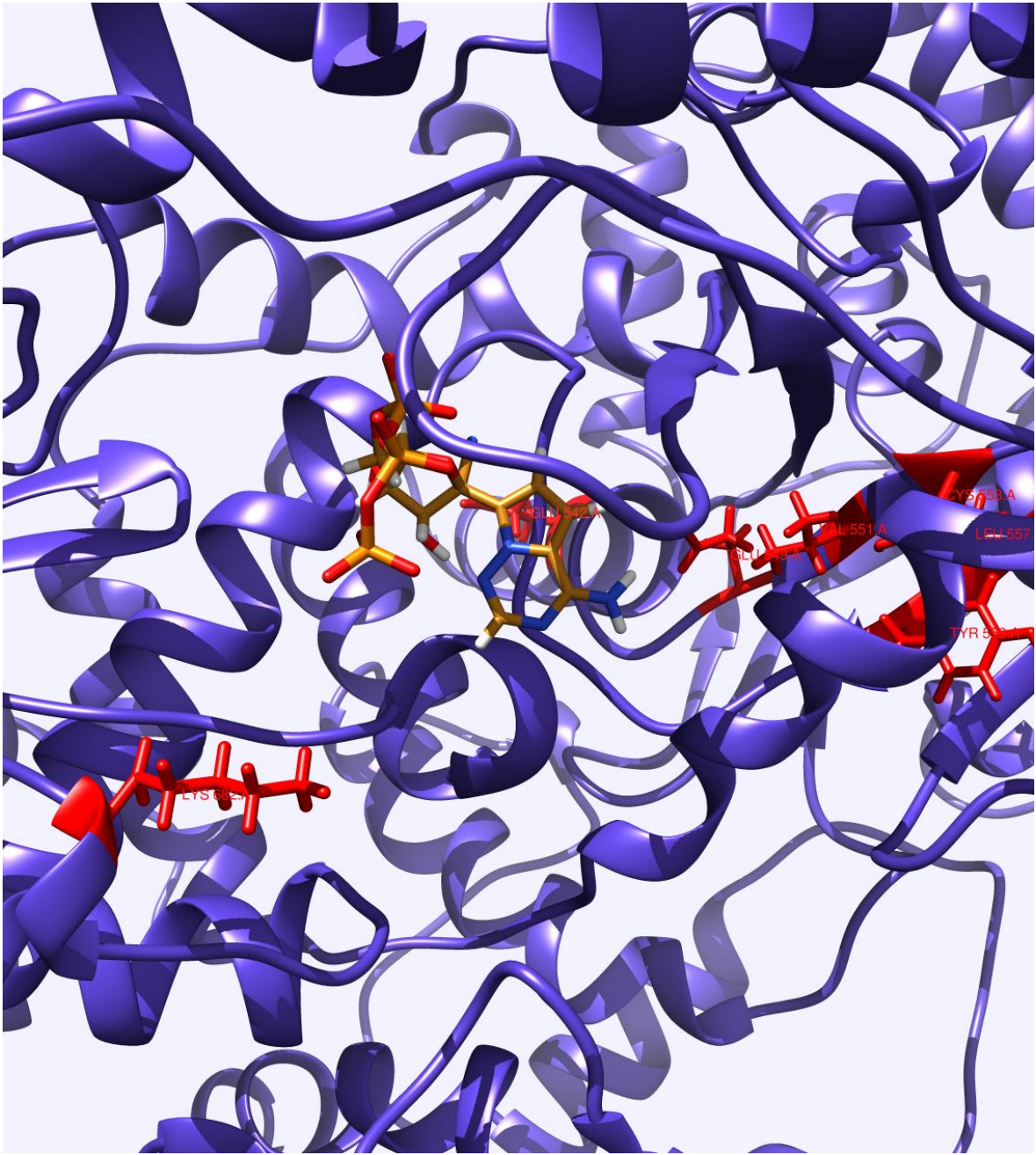

RemTP interaction 229 with NSP-12 SARS-CoV